# Magnetic phase diagram of sigma-phase $Fe_{55}Re_{45}$ compound in the H-T coordinates

Stanisław M. Dubiel[*]

AGH University of Science and Technology, Faculty of Physics and

Applied Computer Science, al. A. Mickiewicza 30, 30-059 Kraków, Poland

Israel Felner and Menahem I. Tsindlekht

Racah Institute of Physics, The Hebrew University,

Jerusalem, Israel 91904

## Abstract

In-field DC and AC magnetization measurements were carried out on a sigma-phase $Fe_{55}Re_{45}$ intermetallic compound aimed at determination of the magnetic phase diagram in the H-T plane. Field cooled, $M_{FC}$, and zero-field cooled, $M_{ZFC}$, DC magnetization curves were measured in the magnetic field, H, up to 1200 Oe. AC magnetic susceptibility measurements were carried out at a constant frequency of 1465 Hz under DC fields up to H=500 Oe. The obtained results provide evidences for re-entrant magnetism in the investigated sample. The magnetic phase diagrams in the H-T plane have been outlined based on characteristic temperatures determined from the DC and AC measurements. The phase diagrams are similar yet not identical. The main difference is that in the DC diagram constructed there are two cross-over transitions within the strong-irreversibility spin-glass state, whereas in the AC susceptibility based diagram only one transition is observed. The border lines (irreversibility, cross-over) can be described in terms of the power laws.

*Corresponding author: Stanislaw.Dubiel@fis.agh.edu.pl

# 1. Introduction

The sigma phase ($\sigma$) is one of numerous examples of the so-called Frank-Kasper (FK) phases, also known as topologically close-paced structures. The $\sigma$ phase in the Fe-Cr system discovered in 1927 [1] has been regarded as the prototype and its crystallographic structure was identified about 25 years later [2]. Its tetragonal unit cell hosts 30 atoms distributed over 5 lattice sites having high coordination numbers (12-15), a characteristic feature of the FK phases. Sigma-phase can be formed only in alloys in which at least one element is a transition metal. In the simplest case i.e. binary alloys, there are known 43 cases in which the occurrence of $\sigma$ has been reported [3]. An interest in $\sigma$ can be classified into two categories: (1) industrial and (2) scientific. The former stems from the fact that it often precipitates (as an extra phase) in technologically important materials, for example, ferritic and/or martensitic steels or super alloys. The presence of $\sigma$ in these materials, even in small percentage, causes a serious deterioration of their useful properties including corrosion resistance, creep strength, impact toughness or tensile ductility. In other words, its presence in these materials is highly undesired. On the other hand, efforts has been recently undertaken to take advantage of its high hardness, in order to strengthen materials properties [4]. Scientifically, $\sigma$ has been of interest *per se*, due to its complex crystallographic structure which is further complicated by lack of stoichiometry. In fact, for a given alloy, $\sigma$ exists in a certain range of composition. Consequently, its physical properties can be tailored by changing the composition. Magnetic properties turned out to be very sensitive to the chemical composition. In the $Fe_{100-x}V_x$ system, in particular, the Curie temperature can varied between few Kelvin and more than 300 K [5]. Regarding the magnetism of $\sigma$, the subject of the present study, it has been so far found in four binary Fe-based alloys viz. Fe-V [5-6], Fe-Cr [7], Fe-Re [8] and Fe-Mo [9]. Initially, the magnetism of $\sigma$ in Fe-Cr and Fe-V alloys was regarded as ferromagnetism (FM). However, recent studies indicate more complex magnetic structure than initially anticipated. It appears that, for both alloys a re-entrant character of the magnetism has been evidenced [11,12]. Also akin character of magnetism has been revealed for $\sigma$ in Fe-Re [8] and Fe-Mo [9,13] alloys.

The aim of the present study is to shed more light on the magnetism of $\sigma$ in the Fe-Re system. The previous study performed on a series of $\sigma$-$Fe_{100}Re_x$ (x=43-53) alloys showed that the Curie temperature, $T_C$, strongly depends on the composition and varies between ~65 K for x=43 and ~23 K for x=53 [8]. At lower temperatures, a transition into a spin-glass (SG) state was detected. The SG state was found to be magnetically heterogeneous i.e. weak and strong irreversibility sub-states were identified. A transition from FM to SG in re-entrant SGs takes place at the so-called irreversibility temperature, $T_{ir}$, while a second transition occurs at the so-called cross-over temperature, $T_{co}$. According to the mean-field theory these transitions also occur in external DC magnetic fields, H. The *loci* of $T_C$, $T_{ir}$ and $T_{co}$ in the H-T coordinates constitute the so-called H-T magnetic phase diagram. Recently, based on in-field DC and AC magnetizations measurements, we constructed such phase diagram for a σ-$Fe_{47}Re_{53}$ sample [14]. Here we report on the H-T phase diagram for a σ-$Fe_{55}Re_{45}$ compound and discuss it in terms of the Gabay-Thoulouse model [15].

## 2. Experimental

### 2.1. Sample

### 2.1. The sample

The σ-$Fe_{55}Re_{45}$ sample was prepared in the following way: powders of elemental iron (3N+ purity) and rhenium (4N purity) were mixed in suitable proportions and masses (2 g), and subsequently pressed into a pellet. The pellet was melted in an arc furnace under protective atmosphere of argon. The resulted ingot was re-melted three times in order to improve chemical homogeneity. The ingot was next annealed at 1803 K for 5 hours in vacuum, and, ultimately, quenched into liquid nitrogen. The loss of mass of the fabricated sample was below 0.01% of its initial value, hence the nominal composition can be regarded as the real one. Powder X-ray diffraction patterns at room temperature confirm the tetragonal structure. More details on structural and electronic properties of this sample can be found in Ref. [16].

### 2.2. DC and AC magnetization measurements

DC Magnetization (M) measurements at various applied magnetic fields (H) in the temperature interval 5 K < T < 80 K, have been performed by a commercial (Quantum Design) superconducting quantum interference device (SQUID) magnetometer with sample mounted in gel-cap. Prior to recording the zero-field-cooled (ZFC) curves, the SQUID magnetometer was always adjusted to be in a *"real"* H = 0 state. The temperature dependence of the FC and ZFC branches were taken via warming the samples. The real ($\chi'$) and imaginary ($\chi''$) AC magnetic susceptibilities, at $H$ ranging up to 500 Oe, were recorded with a home-made pickup coil method inserted into this SQUID at an amplitude of $h_0$=0.05 Oe at the frequency of 1465 Hz.

## 3. Results and Discussion
### 3.1. DC measurements

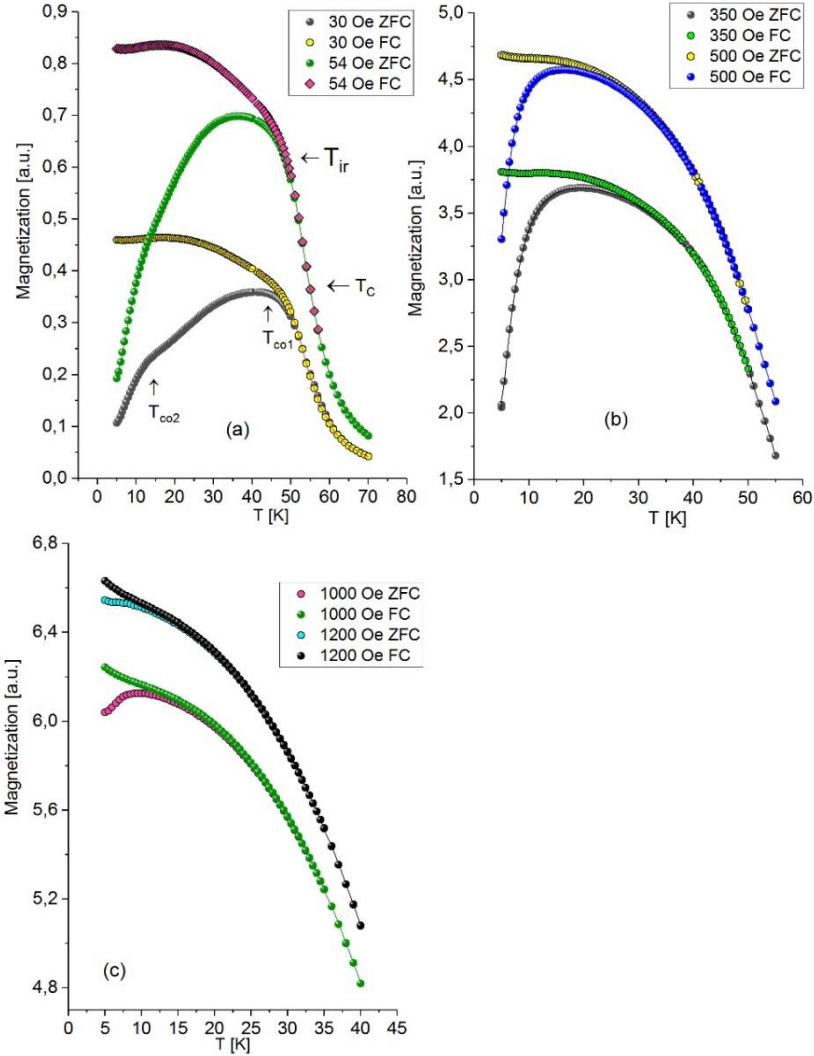

Fig. 1 Selected ZFC and FC magnetization curves vs. temperature and applied magnetic field as displayed in legends. The arrows indicate positions of the characteristic temperatures.

### 3.1.1. Characteristic temperatures

Going from high to low temperatures (Fig. 1), four characteristic temperatures can be distinguished in the $M_{FC}$ and $M_{ZFC}$ curves (1) $T_C$ defined by the inflection point and interpreted as the magnetic ordering (Curie) temperature, (2) $T_{ir}$ defined by the bifurcation of the two curves, and associated with a transition into a spin-glass (SG) state, (3) $T_{co}$ defined by the maximum value of $M_{ZFC}$ curve, and interpreted as a cross-over from a weak into a strong irreversibility state of SG, and (4) $T_{co2}$ defined by a "knee" in the $M_{ZFC}$ curve observe at $T < T_{co}$, which can be interpreted as a transition from the strong irreversibility state into a second stronger one. All these temperatures, except $T_{ir}$, can be precisely determined from the temperature dependence of $dM_{ZFC}/dT$, as shown in Fig. 2.

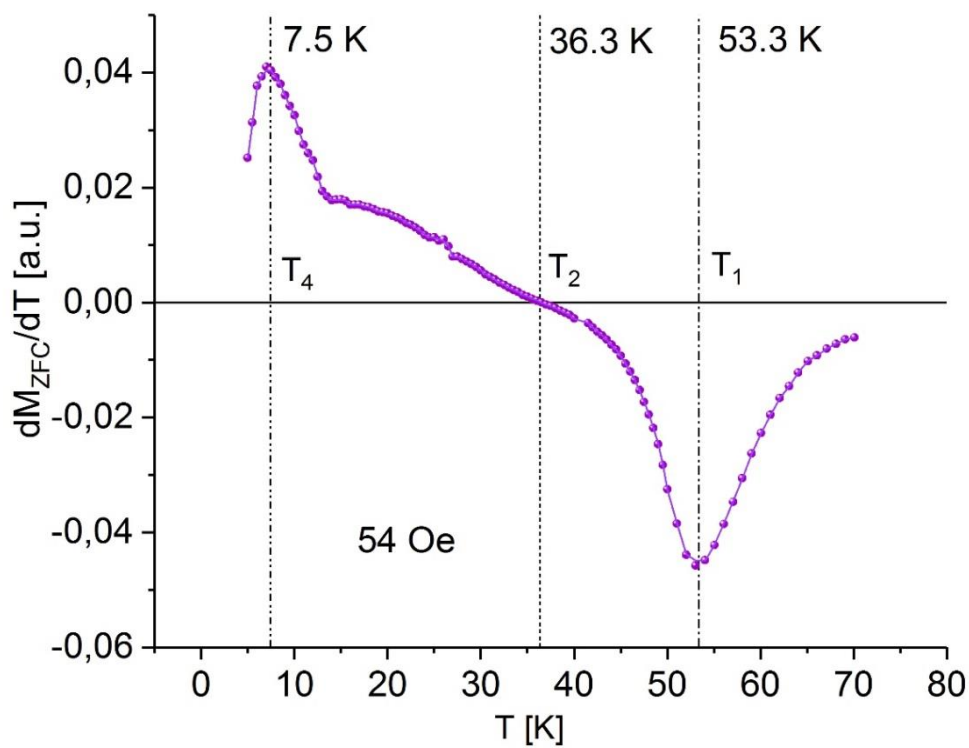

Fig. 2 The temperature dependence of the first temperature derivative of the ZFC-magnetization curve measured at 54 Oe. Characteristic temperatures $T_1=T_C$, $T_2=T_{co1}$ and $T_4=T_{co2}$ and their values are marked.

Generally speaking, well above $T_C$ all M(T) branches at various applied fields, exhibit the typical PM shapes and adhere closely to the Curie-Weiss (CW) law:

$$\chi(T) = \chi_0 + C/(T-\theta) \qquad (1)$$

where $M(T)/H=\chi(T)$, $\chi_0$ is the temperature independent part of the susceptibility $\chi$ (M/H), $C$ is the Curie constant, and $\theta$ is the CW temperature. As shown in Fig. 3, the least squared fit of linear $1/\chi$ versus T plot yields: $C$=1.44 emu K/mol Oe and $\theta$ = 58 K. The PM effective moment ($P_{eff}$) deduced from $C$ for metallic Fe is 4.6 $\mu_B$/Fe, a value which is a bit smaller than $P_{eff}$ of divalent or trivalent Fe ions. The obtained positive $\theta$ agrees well with the $T_C$ value discussed above indicating an isotropic FM state.

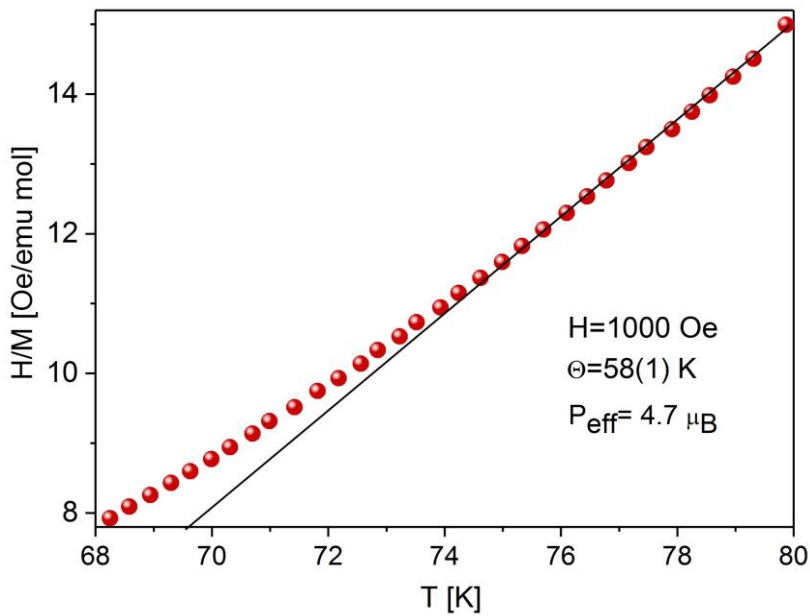

Fig. 3 Reciprocal DC susceptibility, $1/\chi$, vs. temperature, T, for the sample measured in the applied field of 1000 Oe. The solid line is the best fit of the linear data.

Determination of $T_{ir}$ is trickier. The most frequently used way is to plot the difference between the $M_{FC}$ and $M_{ZFC}$ curves vs. temperature and determine the temperature below which the difference is positive. Examples of such obtained results for various H, are displayed in Fig. 4.

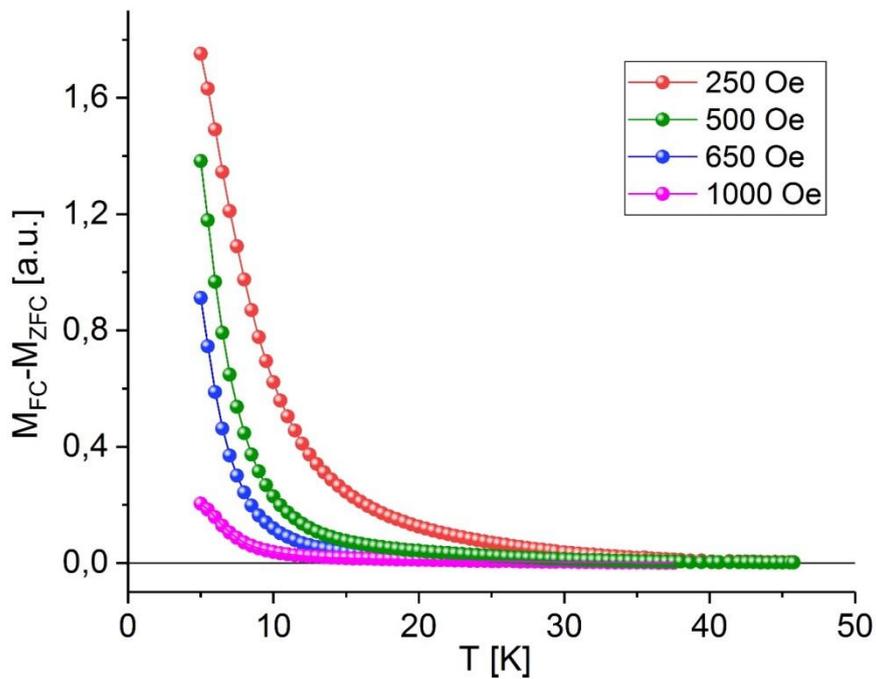

Fig. 4 Difference between the $M_{FC}$ and $M_{ZFC}$ curves vs. temperature, T, for selected values of the applied magnetic field.

### 3.1.2. Degree of irreversibility

The M(T)-curves shown in Fig. 1 give clear evidence that the degree of irreversibility of the SG state, depends both on temperature and on H. It is also clear that with increasing H the difference between the $M_{FC}$ and $M_{ZFC}$ curves decreases. The latter can be obviously related to a decrease of a spin canting, caused by the applied magnetic field. Quantitatively, the irreversibility can be regarded as the difference in areas under the $M_{FC}(T,H)$ and $M_{ZFC}(T,H)$ curves, $\Delta M(H)$:

$$\Delta M(H) = \int_{T_1}^{T_2} (M_{FC}(H) - M_{ZFC}(H)) dT \qquad (2)$$

Where $T_1$=5 K and $T_2$=$T_{ir}$.

The $\Delta$M(H)-values obtained from eq.(2) are presented in Fig. 5. It is evident that the behavior is not monotonic: up to ~200 Oe a rapid increase is seen, followed by an exponential decrease. For comparison, similar data determined for the $\sigma$-$Fe_{47}Re_{53}$ sample [14] have been added. It is obvious that the concentration of Re dramatically affects the degree of irreversibility.

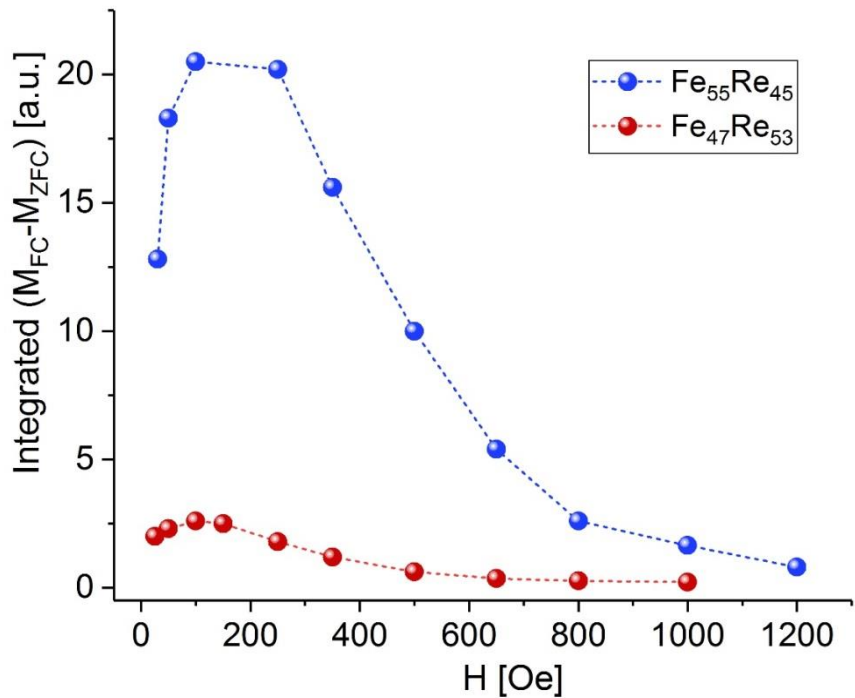

Fig. 5 $\Delta$M(H) for the studied sample as a function of the applied magnetic field, H. For comparison, the corresponding data obtained for $\sigma$-$Fe_{47}Re_{53}$ [14] have been added. Dash lines are the guide to the eye.

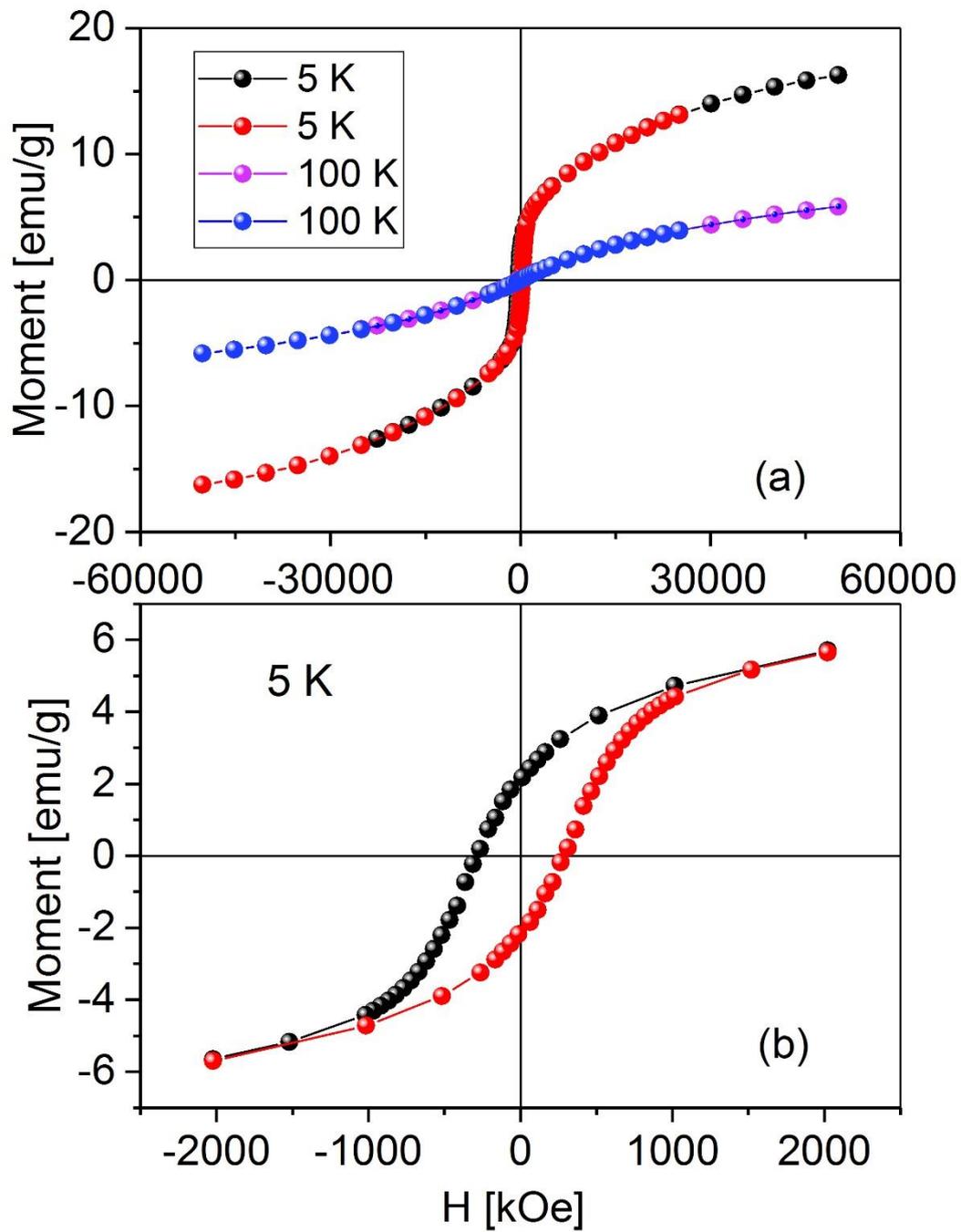

Fig. 6. Isothermal M(H) plots measured (a) at 5 and 100 K, and (b) the hysteresis loop at 5 K.

The isothermal magnetization M(H) curves at 5 K and at 100 K (in the PM state) are shown in Fig. 6 (a). It is readily observed that at 5 K M(H) first increases sharply up to 7 kOe and then tend to saturate, but saturation is not achieved up to 50 kOe. A symmetric hysteresis loop with a coercive field $H_c$= 280 (10) Oe is also shown in Fig. 6(b). On the other hand at 100 K (in the PM range) the M(H) curve is not completely linear as expected but composed of two linear slopes. This M(H) can be fitted as:

$$M(H) = M_s + \chi H, \qquad (3)$$

where $M_s$ is the intrinsic saturation moment due to a FM extra phase and $\chi H$ is the linear PM contribution of the major fraction. $M_s$ obtained is 1.94 emu/g which corresponds to 0.021 $\mu_B$/Fe, a value which is three order of magnitude lower than the PM $P_{eff}$ stated above (Fig. 6***). This FM extra phase is probably due to a tiny undissolved Fe particles remain in the Fe-Re matrix. It should be added, that the coercive field $H_c$=80(10) Oe at 100 K, is much smaller than that shown in Fig. 6.

### 3.1.3. Magnetic phase diagram in the H-T plane

The magnetic phase diagram of $Fe_{55}Re_{45}$ as deduced from the DC magnetization measurements is presented in Fig. 7. It clearly testifies to a re-entrant character of magnetism in this sample, namely on lowering the temperature there is a transition from a paramagnetic (PM) to an intermediate ferromagnetic (FM) and to a spin-glass (SG) states. As expected, the phase transition of the latter state significantly increases with H. The SG state can be divided into two sub states: (1) with a weak irreversibility (SG1) and with a strong one (SG2). The phase field of SG1 has the largest temperature range at 500 Oe. At this field the characteristic lines defined by the *loci* of $T_{ir}(H)$ and of $T_{co}(H)$ show anomalies i.e. there is a change of the dependence on H (note the different colors in Fig. 7).

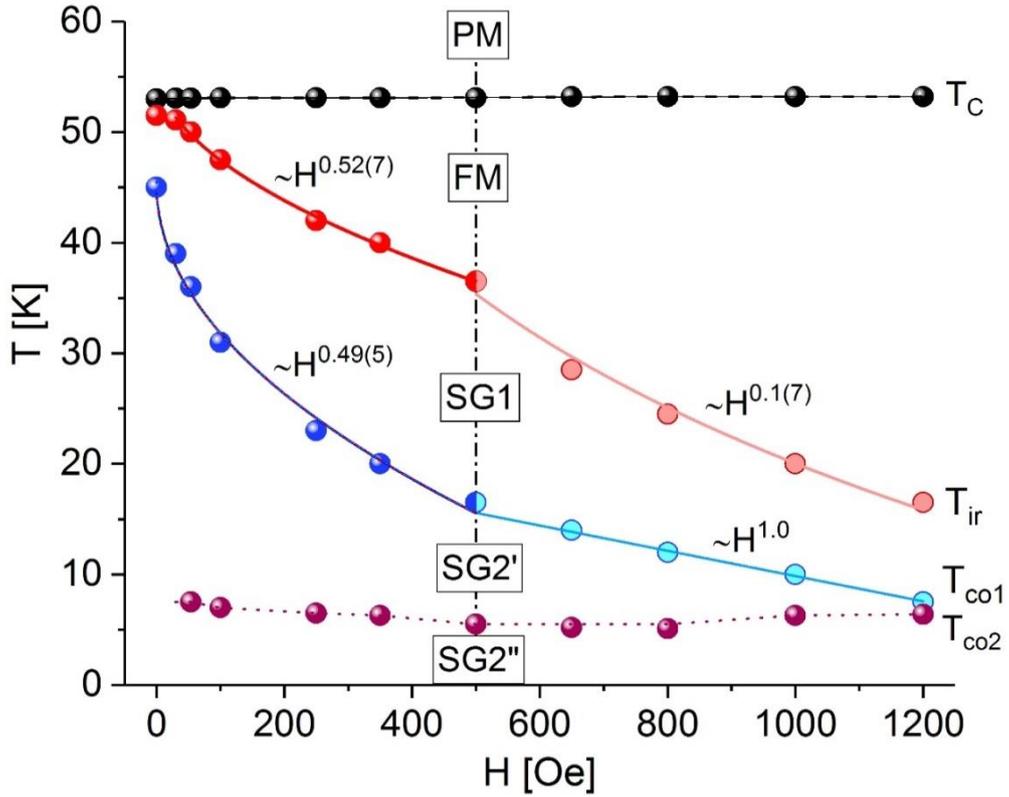

Fig. 7 Magnetic phase diagram of σ Fe$_{55}$Re$_{45}$ as deduced from the DC in-field magnetization measurements. The meaning of the symbols and acronyms can be found in the text. Broken lines are drawn to guide the eye.

$T_{ir}(H)$ and $T_{co}(H)$ dependences can be next used for validation of different predictions relevant to the issue. According to the mean-field theory (MFT), the T-H relationship is:

$$T(H) = T(0) - a \cdot H^{\varphi} \qquad (4)$$

Where T can be either $T_{ir}$ or $T_{co}$.

Three different predictions of $\varphi$ and thus on the T-H relationships can be found throughout the literature as far as the re-entrant SGs states are concerned: (1) $\varphi = 2/3$ (for $T_{ir}$), (2) $\varphi = 2$ (for $T_{co}$) [15] and (3) $\varphi = 1$ [17]. The $T_{ir}$ and $T_{co}$ data shown in Fig. 7 were fitted to eq. (4), and the best fits are indicated by solid lines. Neither $T_{ir}(H)$ nor $T_{co}(H)$ could have been fitted with one value of $\varphi$ within the whole range of H. Furthermore, the obtained values of $\varphi$ disagree with any of the predicted ones, except $\varphi=1$ found for $T_{co}$ in the range $500 \leq H \leq 1200$ Oe. In other words, there is only a qualitative agreement with the Gabay-Thoulouse model [15] i.e. (i) there are two characteristic lines within the SG state, and (ii) both of them decrease with H with two different $\varphi$-exponents. An unusual feature revealed in this study is the existence of a second cross-over transition within the strong-irreversibility SG2 state at $T_{co2} \approx 7$ K which divides it into two sub states, SG2' and SG2". To our best-knowledge, neither such transition has been predicted nor reported so far. The line associated with this transition shows a weak field dependence: up to H=500 Oe a weak decrease and above a weak increase can be seen. At H=1200 Oe this line merge with the "true" strong-irreversibility line i.e. $T_{co1}$.

**3.2. AC measurements**

**3.2.1. Characteristic temperatures**

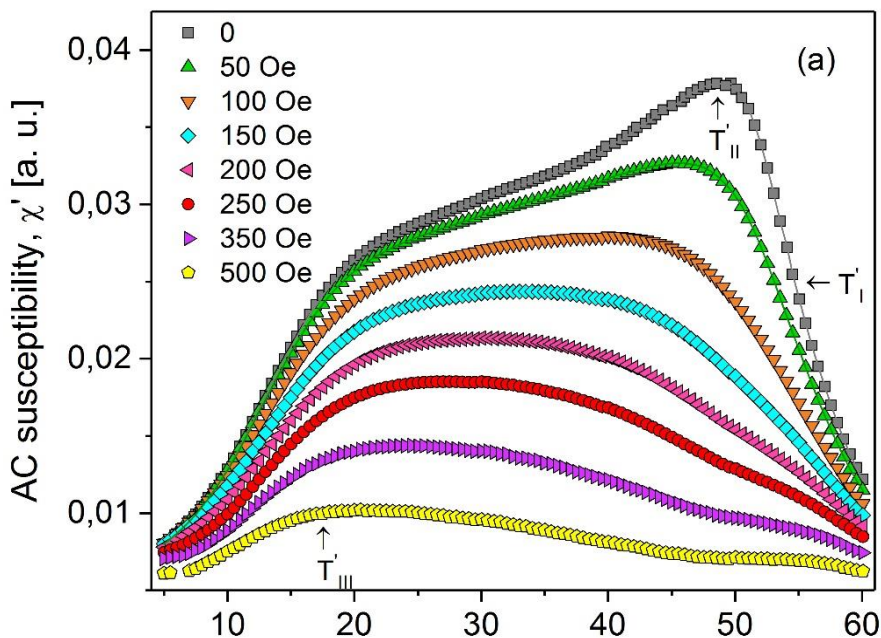

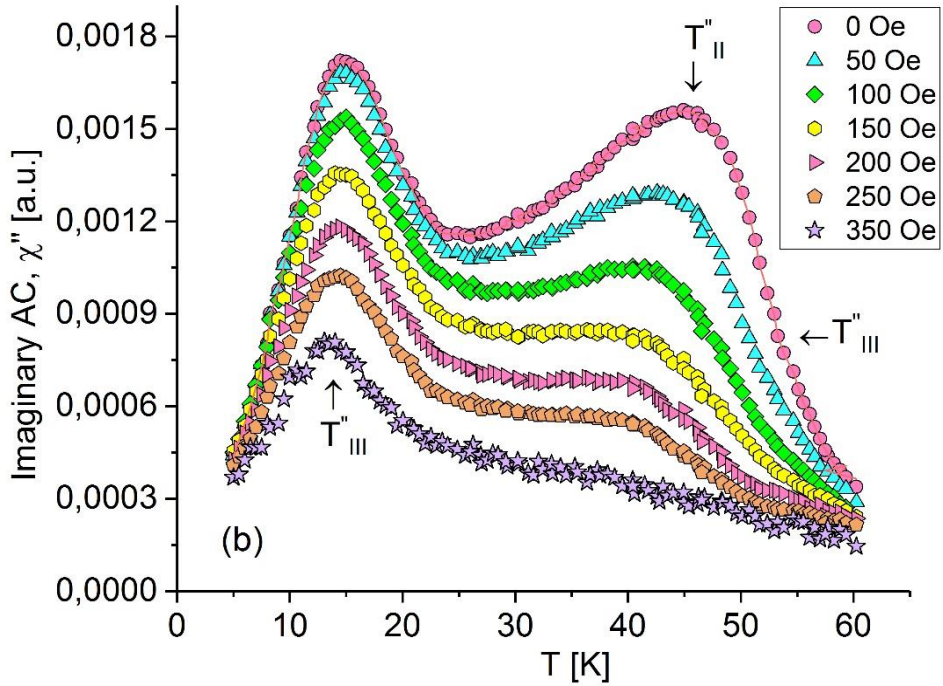

Fig. 8 (a) Real and (b) imaginary parts of the AC susceptibility for sigma-$Fe_{55}Re_{45}$ sample as measured at different values of the applied magnetic field. Characteristic temperatures are indicated by arrows

Three characteristic temperatures $T'_I$, $T'_{II}$ and $T'_{III}$ can be determined both from real ($\chi'$) and as well as from imaginary ($\chi''$) parts ($T''_I$, $T''_{II}$ and $T''_{III}$) of the AC magnetic susceptibility curves. $T_I$ from the inflection point in $\chi'$, hence corresponding to the Curie temperature, $T_C$, $T_{II}$ from the maximum in $\chi'$ and from the first maximum in $\chi''$, and $T_{II}$ from the "knee" in $\chi'$ and from the second maximum in $\chi''$. Approximate positions of these temperatures are marked in Fig. 8 while Fig. 9 illustrates the way of their determination. The temperature $T_{II}$ is usually associated with the spin-freezing temperature, and $T_{III}$ can be interpreted as the cross-over temperature from the weak into the strong irreversibility SG state.

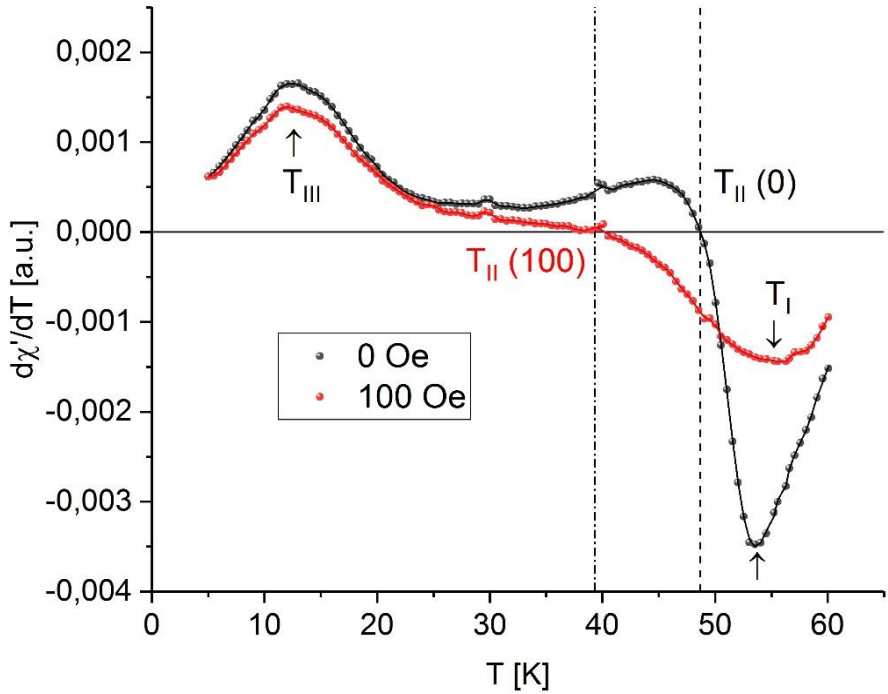

Fig. 9 First temperature derivative of the real part of the AC susceptibility, d$\chi$'/dT, vs. temperature for obtained from the $\chi$'(T) curves measured at 0 and 100 Oe. Positions of the characteristic temperatures are indicated by arrows.

### 3.2.2. Magnetic phase diagram

The magnetic field dependences of these three temperatures are displayed in Fig. 10 yielding the corresponding magnetic phase diagram in the H-T plane. It shows three magnetic phase fields viz. FM, SG1 and SG2 whereas in the phase diagram based on the DC measurements (Fig. 7) there are four magnetic phase fields viz. FM, SG1, SG2' and SG2". The difference stems from the fact that the AC susceptibility curves do not trace the irreversibility transitions at $T_{ir}$. Consequently, the FM phase field in the AC magnetic phase diagram is wider than the one in the DC magnetic phase diagram. Noteworthy, the value of $\varphi=0.7(1)$ for $T_{II}$ (the maximum in the $\chi$') observed in the $T_{II}$(H) line for H $\leq$ 250 Oe agrees well with $\varphi=2/3$ predicted for the cross-over line corresponding to spontaneous breaking of the replica symmetry [15].

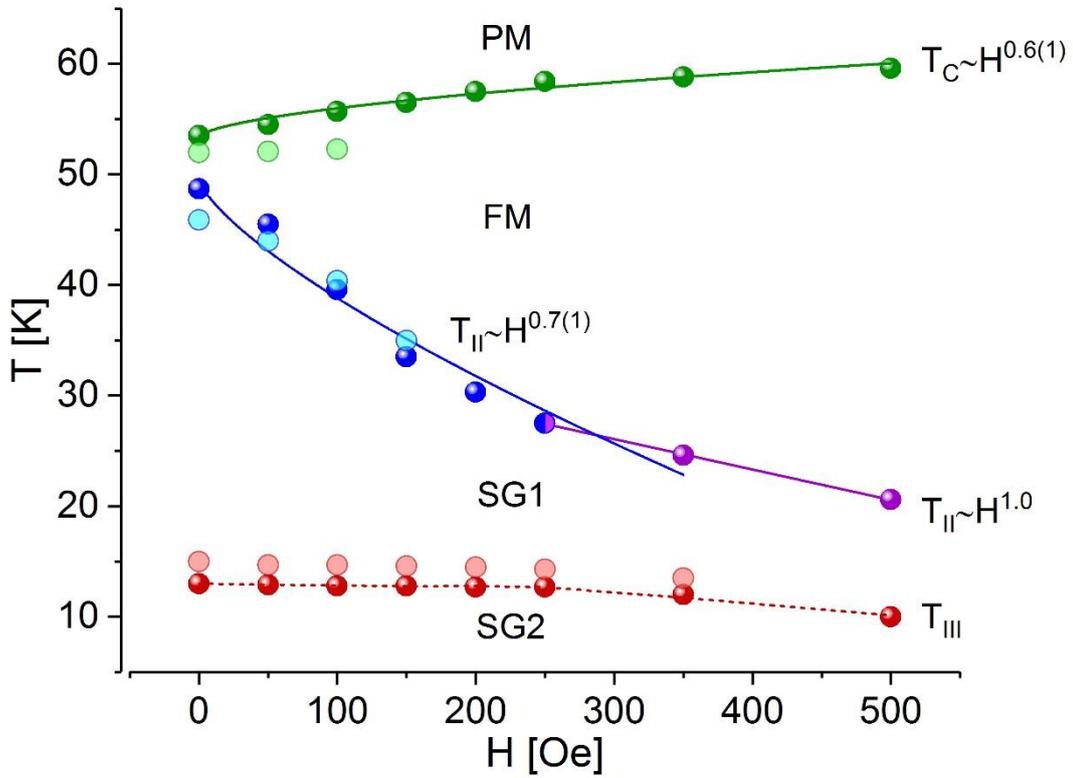

Fig. 10 Magnetic phase diagram in the H-T plane of sigma $Fe_{55}Re_{45}$ sample deduced from AC susceptibility measurements. Full and open symbols were derived from the real imaginary parts of the susceptibility respectively. The solid lines represent the best fits to the data in terms of eq. (4).

## 4. Discussion

The re-entrant character of magnetism viz. PM→FM→SG of $\sigma$-$Fe_{55}Re_{45}$ studied sample was revealed from DC magnetization measurements performed at 100 Oe [8]. The present measurements show that this re-entrance persists up to 1200 Oe. There is a general qualitative agreement with the Gabay-Thouless (GT) model i.e. the transition line from the FM into the SG state (irreversibility line) decreases with H in line with a power law line but with different exponents, $\varphi$. In addition, it shows a cross-over at 500 Oe (value of $\varphi$ changes). As predicts the GT model, there is also a second transition line, known as the cross-over line, marking a transition from a weak into a strong irreversibility SG states, which also decreases with H, but with a different value of $\varphi$ than expected from the model. Furthermore, at 500 Oe, $\varphi$ changes from ~0.5 to 1.0. Noteworthy, $\varphi=1.0$ agrees with the prediction by Dubiel et al. [17]. However, the most significant departure from the GT prediction for the re-entrant SGs, is the existence of a third line (almost independent of H) within the strong-irreversibility SG state. Its existence has been found from both DC and as well as from AC measurements. One may ask a question on whether the presently new observation can be used to validate the GT model. If the answer is yes, then the model fails because: (i) the extracted values of $\varphi$ do not agree with the predicted ones. In addition, (ii) the strong-irreversibility SG state studied here is heterogeneous i.e. there are two cross-over lines, in contrast to the GT model which predicts only one such line.

However, another question may be raised on whether the GT model can be applied to such a complex system like the sigma phase. Its complex crystallographic structure (five different lattice sites with different coordination numbers and distances to the nearest-neighbors) obviously leads to a complex magnetic structure as well. Consequently, there are, at least, five different magnetic sites i.e. Fe atoms occupying these sites have different magnitudes of magnetic moments, different anisotropy and possibly different orientations. Indeed, theoretical calculations performed for $\sigma$-FeRe [8] and $\sigma$ in other Fe-based alloys e. g. Fe-Cr [18], Fe-Mo [19] predict not only different Fe magnetic moments at the different lattice sites, but also antiferromagntic coupling for some of them. Consequently, the overall coupling between the five magnetic moments may be a complex combination of temperature and applied magnetic field. In other words, the spin relations and conditions in the sigma phase are much more different than those assumed in the GT model [15]. Under these circumstances it is not surprising that the GT model does not agree quantitatively with the experimental data presented here case.

Concerning the phase diagram determined from the AC measurements, it is similar to the DC one (for $H \leq 500$ Oe) except for the absent $T_{ir}$-line. Consequently the FM phase field is wider. The H-dependence of the $T_{II}$-line agrees quantitatively with the prediction of the GT model for $H \leq 250$ Oe and with that of Dubiel et al. for higher fields [ 17].

**5. Summary**

Based on the DC and AC magnetization measurements magnetic phase diagrams of the sigma-phase $Fe_{55}Re_{45}$ compound in the H-T coordinates have been constructed. The obtained phase diagrams have similarities and differences. Both of them testify to the re-entrant character of magnetism of the studied compound, viz. on lowering temperature there is a transition from a paramagnetic (PM) to a spin-glass (SG) state with an intermediate ferromagnetic (FM) phase. However, they differ as far as the border between the FM and SG phases is concerned, as well as on the number of lines within the SG state: two lines from DC measurements but one line deduced from the AC measurements. The line constituted by the *loci* of $T_{ir}(H)$ values in the DC magnetization measurements is not visible in the AC susceptibility measurements. Furthermore, there is a difference in: (i) the $\varphi$ values in the corresponding lines, (ii) in their position of the border of the ground state and in (iii) the Curie temperature dependence on H. The exponents in the $T_{II}(H)$ line agree with the MFT predictions [15,17].

**Acknowledgements**

This work was financed by the Faculty of Physics and Applied Computer Science AGH UST statutory tasks within subsidy of Ministry of Science and Higher Education in Warsaw.